%% file: main.tex
\documentclass{article}

\usepackage{arxiv}

\usepackage[utf8]{inputenc} 
\usepackage[T1]{fontenc}    
\usepackage{hyperref}       
\usepackage{url}            
\usepackage{booktabs}       
\usepackage{amsfonts}       
\usepackage{amssymb, amsmath, mathtools, amsthm} 
\usepackage{float}
\usepackage{nicefrac}       
\usepackage{microtype}      
\usepackage{lipsum}
\usepackage{graphicx}
\usepackage{caption}
\usepackage{subcaption}
\usepackage{multicol}
\usepackage{todonotes}

\usepackage[
backend=biber,
style=numeric,
citestyle=numeric-comp,
sorting = nyt,
maxbibnames = 99
]{biblatex}

\renewbibmacro{in:}{} 
\DeclareFieldFormat{pages}{#1} 

\usepackage{tikz,pgfplots,pgf}
\usetikzlibrary{matrix,shapes,positioning}
\usetikzlibrary{arrows,backgrounds}
\usepgflibrary{shapes.multipart}

\newtheorem{definition}{Definition}

\newtheorem{remark}{Remark}

\title{Grouping of Contracts in Insurance using Neural Networks}

\author{
    Mark Kiermayer \\
  Department of Natural Science\\
  University of Applied Sciences Ruhr-West\\
  \texttt{mark.kiermayer@hs-ruhrwest.de} \\
   \And
    Christian Weiß\\
  Department of Natural Sciences\\
  University of Applied Sciences Ruhr-West\\
  \texttt{christian.weiß@hs-ruhrwest.de} \\
}

\begin{document}
\maketitle

\begin{abstract}
    Despite the high importance of grouping in practice, there exists little research on the respective topic. The present work presents a complete framework for grouping and a novel method to optimize model points. Model points are used to substitute clusters of contracts in an insurance portfolio and thus yield a smaller, computationally less burdensome portfolio. 
    This grouped portfolio is controlled to have similar characteristics as the original portfolio. We provide numerical results for term life insurance and defined contribution plans, which indicate the superiority of our approach compared to $K$-means clustering, a common baseline algorithm for grouping. Lastly, we show that the presented concept can optimize a fixed number of model points for the entire portfolio simultaneously. This eliminates the need for any pre-clustering of the portfolio, e.g. by $K$-means clustering, and therefore presents our method as an entirely new and independent  methodology.
\end{abstract}

\keywords{Grouping\and Term Life Insurance \and Defined Contribution Plan \and Non-linear Optimization \and Neural Networks \and Ensemble \and LSTM  \and $K$-means clustering \and Supervised Learning}

\input{Sections/Section_Intro.tex}
\input{Sections/Section_Grouping.tex}
\input{Sections/Section_Methodology.tex}

\input{Sections/Section_Data.tex}
\input{Sections/Section_Num_Results.tex}
\input{Sections/Section_Conclusion.tex}

\section*{Acknowledgements}
We would like to thank Ralf Korn, who stimulated the extension of our research to pension plans. Furthermore, we gratefully acknowledge many useful comments and suggestions of Hans-Joachim Zwiesler and Anselm Haselhoff during the preparation of the paper.

\printbibliography

\input{Sections/Appendix.tex}

\end{document}

%% file: Sections/Section_Intro.tex
\section{Introduction}

Applying new methods of machine learning to challenge the performance of traditional computational schemes in insurance gets more and more attention in recent years, see e.g. \cite{Aleandri17,RiWue18,Wuet18,DepSchevWuet,DuEtAl}. At the same time, feasibility of algorithms plays a key role, leading to several approaches with the aim to reduce the computational complexity of a task, see e.g. \cite{LSMC_solv_II_part_I,LSMC_solv_II_part_II,Noert,HejJac}. A particularly relevant application in insurance is the grouping of contracts. Grouping aims to detect clusters in a portfolio and replace each cluster by a single representative contract, alias model point. Representativeness is to be interpreted as preserving selected key feature of the original portfolio. The model point is in general not part of the original portfolio. 
This simplification, i.e. the reduction of model points, results in a more homogeneous portfolio and in significant time savings for policy-by-policy based computations. The relevance of grouping and key considerations are  discussed in e.g. \cite{BELV, PenAlex, Noert}. In practice, it is a common choice to apply $K$-means clustering and use its centroids as model points. \\
For model points in grouping, there are two strands of literature, stochastic dominance analysis and constructive, algorithmic methods. In \cite{DenTru}, \citeauthor{DenTru} build on earlier work of \citeauthor{Frostig01}, see \cite{Frostig01}, and derive lower and upper bounds for the risk of a portfolio, measured by its TVaR. The second strand of literature aims to determine the hypothetical portfolio consisting of a reduced number of model points, see e.g. \cite{GofGue15, BlanDorSal17}. In \cite{GofGue15}, the authors use a $K$-means cluster assignment and generalize the respective model points, i.e. centroids,  by transitioning to a weighted sum. How to determine model points is a key question in grouping. Building upon previous work  \cite{Kiermayer19}, we follow the two-step procedure of \cite{GofGue15} and provide further generalization for how to non-linearly optimize model points, based on a formal and novel definition of the task. In particular, we employ neural networks that control for arbitrary risk features of the grouped portfolio. To do so, we intertwine the computation of a risk features and the grouping procedure, which controls for similarity of these respective risk features. Eventually, our approach can even be applied without a $K$-means clustering preprocessing of the original portfolio and can therefore eliminate any dependency on the $K$-means algorithm.  We provide numerical analyses of our methodology for portfolios consisting of term life insurance contracts and defined contribution pension plans, where we control for policy values in the resulting portfolios. The results show vast improvements of a $K$-means baseline grouping. Hence, the present article contributes to the body of literature on machine learning techniques in insurance and the practice of grouping. \\


The remainder of this paper is structured as follows. First, we  formally introduce the task of grouping and respective notations, see Section \ref{sec_grouping}. Next, we outline the general concept of how to optimize model points, see Section \ref{sec_methodology}. Based on modeling assumptions and data summarized in Section \ref{sec_data}, we then provide numerical results for term life insurance and pensions, each controlling for the portfolio's policy values, in Section \ref{sec_num_results}. Here we will analyse the ability of neural networks to replicate the classical, actuarial calculation of policy values of individual contracts, as well as the actual task of how to optimize model points in grouping. Lastly, we summarize the results in Section \ref{sec_conclusion} and provide and mention tasks for further research.

%% file: Sections/Section_Grouping.tex
\section{Concept of Grouping} \label{sec_grouping}
Sound risk management in insurance often requires computationally expensive analysis, e.g. due to regulatory capital requirements by Solvency II. The Solvency II directive demands insurance companies to derive the full loss distribution for calculating the solvency capital requirement (SCR). To calculate the full loss distribution reasonably accurate, several hundred thousand simulations each consisting of at least $1~000$ Monte Carlo simulations would be necessary. In its pure form, this is a challenging yet computationally almost impossible task. One way to reduce the run time of corresponding algorithms is to use approximations of the methodology, as done in \cite{LSMC_solv_II_part_I, LSMC_solv_II_part_II, LeiHoeKletBet13, KraNikKor}. A second, potentially complementing option is to reduce the number of model points of the portfolio which the algorithm is applied to, while preserving certain properties  of the portfolio, e.g. cash flows or risk features. Motivated by the similarity of computationally less expensive properties one expects, that also other, computationally more expensive properties, e.g. SCR, of the actual portfolio and its simplified version are alike. In this work, we focus on the second task of reducing the number of model points, which we refer to as grouping of contracts. Subsequent extensive computations of quantities, as e.g. the SCR, are not part of this article. \\
To begin with, let us formally state the problem. For $K, N\in\mathbb{N}$, $K\leq N$, we introduce 
\begin{align} \label{def_P_KN}
    &\mathcal{P}_{K,N}=\bigg\lbrace\left\lbrace \left(x^{(1)},s^{(1)}\right), \ldots \left(x^{(K)},s^{(K)}\right)\right\rbrace ~\bigg\vert&& \sum_{i=1}^{K}s^{(i)} = N, \nonumber \\
    & &&(x^{(i)},s^{(i)})  \in\mathbb{R}^{n}\!\times\!\mathbb{N},~ i=1,\ldots,K \bigg\rbrace ,
\end{align}
as the family of portfolios with $K$ contracts, resp. model points, $x^{(i)}\in \mathbb{R}^n$ held in corresponding quantities $s^{(i)}\in\mathbb{N}$, summing up to a total number of $N$ contracts. We call $\Tilde{P}\in\mathcal{P}_{K,N}$ a \textit{grouping} of $P\in\mathcal{P}_{H,N}$, if $K<H\leq N$. \\
Note, that we do not explicitly require contracts $x^{(1)}, \ldots,x^{(K)}$ to be distinct, allowing for trivial transformations of e.g. $\lbrace (x^{(1)},s^{(1)}),(x^{(2)},s^{(2)})\rbrace \in \mathcal{P}_{2,s^{(1)}+s^{(2)}}$ to $\lbrace (x^{(1)},s^{(1)}+s^{(2)})\rbrace \in\mathcal{P}_{1,s^{(1)}+s^{(2)}}$, if $x^{(1)}=x^{(2)}$. Also, the definition of $\mathcal{P}_{K,N}$ mirrors the idea of $K$-means clustering, where contract $x^{(i)}$ is the centroid of the $i$-th cluster with $s^{(i)}$ members. We will later use $K$-means clustering, a common choice for grouping, see \cite{Noert, PenAlex}, as a baseline and compare it with our novel approach, which optimizes its centroids. \\
To quantify the quality of the grouping we further define user-selected \textit{risk features} $R_1,\ldots,R_l$, defined by
 $$R_i: \lbrace P\in\mathcal{P}_{K,N}\vert K,N\in\mathbb{N}, K\leq N \rbrace \xrightarrow{} \mathbb{R}^{r_i}, \quad r_i \in\mathbb{N},$$
which, for $P\in\mathcal{P}_{K,N}$, satisfy
\begin{align} \label{R_linearity}
    R_i(P):= \sum_{i=1}^{K}R_i(\lbrace (x^{(i)}, s^{(i)} \rbrace) = \sum_{i=1}^{K}s^{(i)}R_i(\lbrace (x^{(i)}, 1 \rbrace).
\end{align}
On the one hand, the linearity w.r.t. scalar multiplication in \eqref{R_linearity} is a crucial assumption, as it shows how grouping, which results in multiple identical contracts $x^{(i)}$ in a portfolio $P\in\mathcal{P}_{K,N}$, reduces the computational cost when calculating the risk features $R_i$. On the other hand, this assumption is not highly restrictive, as it holds for all actuarial present values, i.e. discounted expected cash flow computations, which lie at the heart of most calculations in life insurance. Note, that we impose linearity of $R_i$ w.r.t. the number of contracts $s^{(i)}$, but not the the components of contract $x^{(i)}$. Exemplary, practical choices for $R_i$ include expected premium payments, profit signatures or policy values. Discrepancies between the actual portfolio $P\in\mathcal{P}_{N,N}$ and its grouped version $\Tilde{P}\in\mathcal{P}_{K,N}$ for a given risk feature $R_i$ are then evaluated w.r.t. an user-defined norm $\lVert \cdot \rVert^{(i)}$, i.e. $\lVert R_i(P)-R_i(\Tilde{P}) \rVert^{(i)}$. This leads to the following definition. \\ 

\begin{definition} \label{def_grouping}
Let $K$ be the target complexity of portfolio $P\in\mathcal{P}_{H,N}$, $K\ll H\leq N$ and $R_1,\ldots, R_l$ the risk features of interest with corresponding, user-selected norms $\lVert \cdot\rVert^{(i)}, i=1,\ldots,l.$ Then the task of \textbf{optimal grouping} equals finding $\Tilde{P}\in\mathcal{P}_{K,N}$ such that
\begin{align}
    \Tilde{P} & = {\arg\min}_{P^{'}\in\mathcal{P}_{K,N}} \sum_{i=1}^{l} \lVert R_i(P^{'})-R_i(P) \rVert^{(i)}.
\end{align}
\end{definition}

\begin{remark}
    Definition \ref{def_grouping} represents a classical clustering with an additional constraint. In general, $K$-means clustering does not include an active control for any risk feature $R_i$ and thus does not match Definition \ref{def_grouping}. $K$-means aims for similar risk features $R_1,\ldots,R_l$ of $P\in\mathcal{P}_{N,N}$ and $P\in\mathcal{P}_{K,N}, K<N$, by homogeneously grouping contracts w.r.t. their respective contract details, e.g. age, duration, sum insured, and a so called dissimilarity measure. Using mean values to obtain centroids $x^{(1)},\ldots,x^{(K)}$ implicitly assumes a linear effect of all contract detail on any risk feature $R_i$, meaning the quality of the grouping highly depends on the homogeneity within each cluster. The active control for $R_i$ in Definition \ref{def_grouping} allows for non-linear effects of contract details when forming an optimized centroid $x^{(i)}$.
\end{remark}
\begin{remark}
    For the sake of simplicity, we denote the risk feature $R$ of a single contract $x\in\mathbb{R}^n$ by $R(x):= R(\lbrace (x,1)\rbrace ).$
\end{remark}

Given a proposed grouping $\Tilde{P}\in\mathcal{P}_{K,N}$, it is then up to the user to set and check acceptable thresholds, e.g. $\tfrac{\lVert R_i(P)-R_i(\Tilde{P}) \rVert^{(i)}}{\lVert  R_i(P) \rVert^{(i)}} <\alpha^{(i)} \in [0,1],~\forall i=1,\ldots,l$, which lead to either acceptance or rejection of $\Tilde{P}$. A common approach for vector-valued risk features $R_i$, e.g. expected cash flows or policy values for various points in time $t=1,\ldots, r_i$, is to evaluate discrepancies componentwise and choose thresholds $\alpha^{(i)}_t$ to be increasing in $t$, see \cite{BELV}. This is motivated by the uncertainty of the outcome, which increases as time progresses.  Lastly, it is important to perform a sensitivity analysis of the results to test for stability of the grouping. Corresponding risk features $R_i(P)$ are often based on specific assumptions, e.g. a fixed interest rate, which limits the generality of the grouping unless backtested. In any case, if we have to reject the proposed grouping, the procedure is to be repeated with altered, less restrictive parameters, e.g. an increased $K$ or a norm postulating a different importance of risk factor $R_i$. 

%% file: Sections/Section_Methodology.tex
\section{Methodology} \label{sec_methodology}

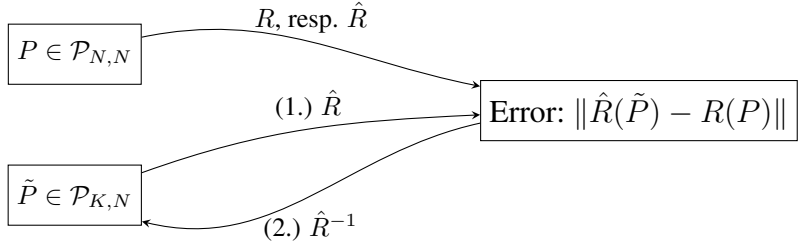
\begin{figure}[htb]
    \centering
    \input{Images/tikz_plots/Sketch_approach.tex}    
    \caption{Illustration of the Methodology of our approach.}
    \label{sketch_approach}
\end{figure}

\begin{figure}[htb]
        \centering
        \input{Images/tikz_plots/Sketch_grouping.tex}    
        \caption{Network Architecture for Grouping of a Cluster of Contracts.}
        \label{fig_sketch_grouping}
    \end{figure}
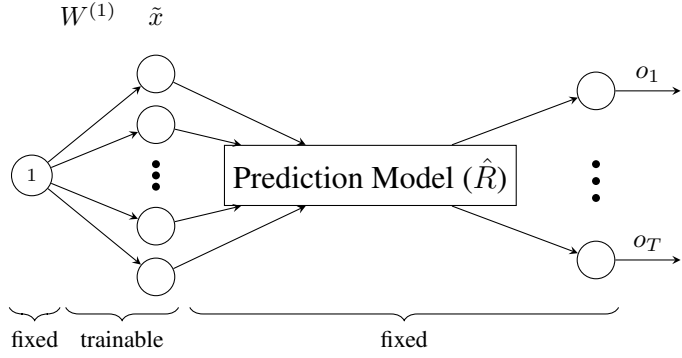

Using previous notation, our approach illustrated in Figure \ref{sketch_approach} is twofold. For grouping portfolio $P\in\mathcal{P}_{N,N}$ our methodology aims to:
\begin{itemize}
    \item[1.] Construct an approximation $\hat{R}~(=(\hat{R_1},\ldots,\hat{R_l}):\lbrace P\in\mathcal{P}_{K,N}\vert K,N\in\mathbb{N}\rbrace\rightarrow \mathbb{R}^{r_1+\ldots r_l})$ of the given risk features of interest $R = (R_1,\ldots,R_l)$, for any individual contract $x\in\mathbb{R}^n$, with $\lbrace(x,1)\rbrace \in P$. Using the assumption of scalar linearity in \eqref{R_linearity}, we cumulatively obtain $R_i(P)$, resp. estimates $\hat{R_i}(P)$, for $i=1,\ldots,l$.
    \item[2.] Based on risk features $R_i(P)$, resp. $\hat{R_i}(P)$, of the actual portfolio $P$, we try to find a grouping $\Tilde{P}\in\mathcal{P}_{K,N}$ with similar properties according to Definition \ref{def_grouping}. Based on a presorting of the portfolio $P$ into $K$ disjoint clusters, we substitute each contract of a cluster $C\subset P$, $C\in\mathcal{P}_{\vert C \vert, \vert C \vert}$ by a single contract $\Tilde{x}$, where $\Tilde{C}=\lbrace (\Tilde{x},\vert C\vert )\rbrace\in\mathcal{P}_{1,\vert C \vert}$ and $\vert C \vert$ the number of contracts in cluster $C$. The substitution $\Tilde{x}$ is chosen, such that it minimizes the function $f_C(\cdot)$ with user-defined norms $\lVert \cdot \rVert^{(i)}$, where
    \begin{align} \label{eq_cluster_error}
        f_C(\Tilde{x}):= \lVert \hat{R}(\Tilde{x})\cdot \vert C \vert-R(C) \rVert := \sum_{i=1}^l \lVert \hat{R}_i(\Tilde{C})-R_i(C) \rVert^{(i)}.
    \end{align}
    It is obvious to see that if there is a perfect match of cluster $C$ and $\Tilde{C}$, i.e. $f_C(\Tilde{x})=0$, it must hold
    \begin{align} \label{eq_cluster_centroid}
        \Tilde{C}\in \hat{R}^{-1}(R(C)),
    \end{align}
    where $\hat{R}^{-1}$ denotes the preimage of function $\hat{R}$. At a slight misuse of notation, as in general $\hat{R}^{-1}(z)$ is not defined for an arbitrary $z\in\mathbb{R}^{r_1+\ldots r_l}$, we denote the procedure of finding an optimal model point by $\hat{R}^{-1}$, see Figure \ref{fig_sketch_grouping}. In Remark \ref{rem_grouping_method} we give further arguments why the notation $\hat{R}^{-1}$ is adequate nevertheless.
\end{itemize}

Computationally, the class of neural networks provides a promising set of models to perform our concept with. Neural networks are generally known to have high approximating capacities, see e.g. \cite{Cybenko}, \cite{Hornik}. On the other hand, neural networks can utilize the preimage of a nested function, i.e. $\hat{R}$, during its training by applying backpropagation. Mention of similar ideas can be found e.g. in \cite{Seletal} and their guided Grad-CAM approach in image classification and detection.\\

\begin{remark}\label{rem_grouping_method}
    \begin{itemize}
        \item[1.] In order to estimate the preimage $\hat{R}^{-1}$ we will apply backpropagation, a supervised and gradient based learning technique for neural networks, and optimize the input $\Tilde{C}$ to $\hat{R}=(\hat{R}_1,\ldots,\hat{R}_l)$ to minimize $f_C(\cdot)$, as defined in \eqref{eq_cluster_error}. This has two main consequences. First, the (numerical) differentiability of $\hat{R}$ is a crucial requirement. The main advantage of working with approximative computations $\hat{R}$, instead of the traditional $R$, is that we can control $\hat{R}$ to be (numerically) differentiable. 
        Secondly, backpropagation is defined for all inputs $\Tilde{C}\in\mathcal{P}_{1,\vert C \vert}$, which means that an approximation of the preimage $\hat{R}^{-1}$ as in \eqref{eq_cluster_centroid} is defined and valid for all target values $R(C)$.

        \item[2.] The function $\hat{R}$ is, in general, not injective, meaning two different contract settings, e.g. $x_1,x_2$, can result in equal risk features, e.g. $\hat{R}(x_1)=\hat{R}(x_2)$. Hence, the optimized model point $\Tilde{C}$ in \eqref{eq_cluster_centroid} may not be unique. However, concerns about the uniqueness of $\Tilde{C}$ can be addressed by including additional constraints for model points, e.g. restricting its age to the range of ages of the respective cluster members. Also, more risk  features $R=(R_1,\ldots,R_l)$ naturally constrain the set of possible solution $\hat{R}^{-1}(R(C))$.
        
        \item[3.] To obtain $\Tilde{C}$, we can either use the traditional quantities $R$, i.e. $\hat{R}^{-1}\left( R(C) \right)$ as done in \eqref{eq_cluster_centroid}, or work with the approximate by proposing $\hat{R}^{-1}\left( \hat{R}(C) \right)$, which implicitly states $\hat{R}$ to be the true target and not an approximation. In any case, we need traditionally computed values $R(x)$ for individual contracts $x\in\mathbb{R}^n$ to supervise the construction of $\hat{R}$. Hence, we work with the traditional computation $R(\cdot)$ for target values.
    \end{itemize}
\end{remark}

The first task is to replicate the computation of target features $Y=R(X)$ using neural networks, where $X\in \mathcal{P}_{1,1}$ represents an arbitrary insurance contract. Similar to \cite{elements_stat_learn}, we explicitly mark the arbitrariness of contract $X$ by capital notation. In Section \ref{sec_num_results} we will choose $R$ to be the standard, actuarial principle for calculating policy values, see \eqref{eq_reserve_TL} and \eqref{eq_reserve_DC}. This task is a standard function approximation, which we phrase as a supervised learning problem,  where we are interested in the function $\hat{R}$, such that we minimize
            $$\lVert \hat{R}(X) - Y\rVert .$$
            
The accuracy of prediction model $\hat{R}$ is of paramount importance, as it affects the optimization schema of our model point, see \eqref{eq_cluster_error}, \eqref{eq_cluster_centroid}. Next, we employ a $K$-means algorithm to preprocess the portfolio $P$ for homogeneity. Conditional on $\hat{R}$, we optimize model point $\Tilde{x}$ for each cluster $C$ by backpropagating the error $\lVert o \cdot \vert C \vert- R(C) \rVert$ of the output $o=\hat{R}(\Tilde{x})$. Hence, we effectively aim to optimize the input of the model $\hat{R}$. \\

To do so, we nest the prediction model $\hat{R}$ in a larger model and use a single, constant input, i.e. of value $1$. This input is connected to a hidden layer, where the nodes represent a single contract $\Tilde{x}$, which then again are input to $\hat{R}$. The constant input allows us to train the model for each cluster $C$ w.r.t. a single data point with target value $\tfrac{R(C)}{\vert C \vert}$ and optimize the input $\Tilde{x}$ to $\hat{R}(\cdot)$. This numerically implements the idea of estimating $\hat{R}^{-1}$ as presented in \eqref{eq_cluster_centroid}. For practical reasons, we omit the usage of bias terms for the first hidden layer. This enables us to use $K$-means centroid $\Tilde{c}$ of cluster $C$, as the initial model point $\Tilde{x}$, e.g. by setting $W^{(1)}=f^{-1}(\Tilde{c})$, for an invertible activation function $f$ in the first layer of the network depicted in Figure \ref{fig_sketch_grouping}. For more detail on neural networks we refer to \cite{HaDeBeJe, GoBeCo}. Numeric results for term life contracts and defined contribution plans are provided in Section \ref{sec_num_results}.

\begin{remark}
    The architecture in Figure \ref{fig_sketch_grouping} shows how to substitute one cluster $C$ by a single model point $\Tilde{x}$, in according frequency $\vert C\vert$. Alternatively, we can enrich the presented architecture by allowing for multiple model points, e.g. $\Tilde{x}^{(1)},\ldots,\Tilde{x}^{(m)}$, which are input to the prediction model $\hat{R}$ and create predictions $\hat{R}(\Tilde{x}^{(1)}),\ldots,\hat{R}(\Tilde{x}^{(m)})$. These predictions, i.e. the importance of each model point in cluster $C$, then have to be weighted to obtain a collective output $o=(o_1,\ldots,o_T)$. In Section \ref{sec_num_results}, we provide results for this generalization for defined contribution pension plans and equal weights for all model points of a single cluster $C$, i.e. $o=\tfrac{1}{m}\sum_{i=1}^{m}\hat{R}(\Tilde{x}^{(i)})$. Effectively, this adjustment enables the method to work without a $K$-mean preprocessing of the portfolio, if we set $C=P$.
\end{remark}

%% file: Images/tikz_plots/Sketch_approach.tex
\tikzset{%
  every neuron/.style={
    circle,
    draw,
    minimum size=1cm
  },
  neuron missing/.style={
    draw=none, 
    scale=2,
    text height=0.333cm,
    execute at begin node=\color{black}$\vdots$
  },
}

    \begin{tikzpicture}[x=1.5cm, y=1.5cm, >=stealth]
        \tikzstyle{rectangle_style}=[rectangle, draw, minimum size = 0.8cm]
        
        \draw node at (0, 0.5) [rectangle_style] (portfolio_0) {$P\in\mathcal{P}_{N,N}$};

        \draw node at (5, 0) [rectangle_style] (features) {\large{\text{Error:}} $\lVert \hat{R}(\Tilde{P})-R(P) \rVert$};
        
        \tikzstyle{rectangle_style}=[rectangle, draw, minimum size = 0.8cm]
        
        \draw node at (0, -0.75) [rectangle_style] (portfolio_1) {$\Tilde{P}\in\mathcal{P}_{K,N}$};
        
        \draw[->] (portfolio_0) to [bend left=80, out = 20, in = 177] node [align=center, above] {$R$, resp. $\hat{R}$} (features);
        \draw[->] (portfolio_1) to [bend left=80, out = 10, in = 173] node [align=center, above] {(1.) $\hat{R}$} (features);
        
        \draw[->] (features)[left] to [bend left, out = -4] node [align=center, below = 0.05] {(2.) $\hat{R}^{-1}$} (portfolio_1);
    \end{tikzpicture}

%% file: Images/tikz_plots/Sketch_grouping.tex
\tikzset{%
  every neuron/.style={
    circle,
    draw,
    minimum size=0.5cm
  },
  neuron missing/.style={
    draw=none, 
    scale=2,
    text height=0.333cm,
    execute at begin node=\color{black}$\vdots$
  },
}

    \begin{tikzpicture}[x=1.5cm, y=1.5cm, >=stealth]
        \tikzstyle{rectangle_style}=[rectangle, draw, minimum size = 0.8cm]
        
        \begin{scope}[black, decoration={brace,amplitude=5pt,raise=3pt}]
            \draw[decorate, decoration={mirror}] (-0.2, -0.35) -- (0.25, -0.35) node [black,midway,yshift=-15pt] {\footnotesize
fixed};
            \draw[decorate, decoration={mirror}] (0.3, -0.35) -- (1.3, -0.35) node [black,midway,yshift=-15pt] {\footnotesize
trainable};
            \draw[decorate, decoration={mirror}] (1.4, -0.35) -- (5.2, -0.35) node [black,midway,yshift=-15pt] {\footnotesize
fixed};
        \end{scope}

        \foreach \m/\l [count=\y] in {1}
         \node [every neuron/.try, neuron \m/.try] (input) at (0,0.75) {\scriptsize{$1$}};
        

        \foreach \m/\l [count=\y] in {1,2,4,5}
          \node [every neuron/.try, neuron \m/.try] (h-\m) at (1.1,2.1-\m*0.45) {};
          
        \foreach \x in {0,1,2} 
            \fill (1.1,2.1-1.35+0.1-\x*0.1) circle (1.5pt);
            
        \node [align=center, above] at (1.1,2) {$\Tilde{x}$};

        \draw node at (3, 0.75) [rectangle_style] (EP) {\large{\text{Prediction Model ($\hat{R}$)}}};
        
        \foreach \m/\l [count=\y] in {1}
          \node [every neuron/.try, neuron \m/.try] (output-\m) at (5,1.5) {};
        \foreach \m/\l [count=\y] in {41}
          \node [every neuron/.try, neuron \m/.try] (output-\m) at (5,0) {};
         \foreach \x in {1,...,3} 
            \fill (5, 1.0 -\x*0.15) circle (1.5pt);  
    

         
         \draw [->] (output-1) -- ++(0.75,0) 
            node [above, midway] {$o_{1}$};
        \draw [->] (output-41) -- ++(0.75,0) 
            node [above, midway] {$o_{T}$};


          \foreach \j in {1,2,4,5}
            \draw [->] (input) -- (h-\j);
        \node [align=center, above] at (0.5, 2) {$W^{(1)}$}; 
        
        \foreach \i in {1,2,4,5}
            \draw [->] (h-\i) -- (EP);
        \foreach \i in {1,41}
            \draw [->] (EP) -- (output-\i);

    \end{tikzpicture}

%% file: Sections/Section_Data.tex
\section{Data} \label{sec_data}

In the following, we describe contract features and assumptions for portfolios of term life insurance contracts and defined contribution pension plans separately. In both cases, the risk feature of interest for our grouping procedure are policy values up to the contract's maturity, resp. the maximum retirement age. For computation of policy values we apply actuarial basics and standard notations, as e.g. $t$-year survival, resp. death, probabilities ${}_t p_x$, resp. ${}_t q_x:= 1-{}_t p_x$, of an individual aged $x$ years, conditional that the individual is still alive. For a more detailed review of actuarial principles we refer to \cite{dickson_hardy_waters_2009, Neuburger19}. \\

\begin{table}[ht]
    \centering
    \begin{subtable}[c]{0.5\textwidth}
        \centering
            \begin{tabular}{l|c|c|c} 
                    \multicolumn{1}{c}{ } & \multicolumn{1}{c}{ Feature } & \multicolumn{1}{c}{ $X_i([0,1]^5)$} &  \multicolumn{1}{c}{ Unit} \\ \hline
                    $X_1 $ &Age (current) & $\lbrace 25,\ldots,107\rbrace$ & Years\\ \hline
                    $X_2$ & Sum Insured & $\lbrace 10^3,\ldots,10^6\rbrace$  & Euro\\ \hline
                    $X_3$  & Duration & $\lbrace 2,\ldots,40\rbrace$ & Years \\ \hline
                    $X_4$ & Lapsed Duration  & $\lbrace 0,\ldots,39\rbrace$ & Years  \\ \hline
                    $X_5$ & Interest Rate  & $\left[ -0.01, 0.04 \right]$ & Numeric  \\ \hline
            \end{tabular}
            \subcaption{Term Life Insurance.}
            \label{table_features_values_TL}
    \end{subtable}%
    \begin{subtable}[c]{0.5\textwidth}
        \centering
        \begin{tabular}{l|c|c|c}
                \multicolumn{1}{c}{ } & \multicolumn{1}{c}{ Feature } & \multicolumn{1}{c}{ $X_i([0,1]^5)$}&  \multicolumn{1}{c}{ Unit}   \\ \hline
                $X_1$ & Age (current) & $\lbrace 25, \ldots,60 \rbrace$& Years   \\ \hline
                $X_2$ & Fund Volume & $\left[ 0, 200~000 \right]$& Euro   \\ \hline
                $X_3$ & Salary & $\left[ 20~000, 200~000 \right]$ & Euro\\ \hline
                $X_4$ & Salary Scale  & $\left[ 0.01, 0.05 \right]$  & Numeric\\ \hline
                $X_5$ & Contribution   & $\left[ 0.01, 0.1 \right]$ & Numeric \\ \hline
        \end{tabular}
        \subcaption{Defined Contribution Plan.}
        \label{table_features_pension}
    \end{subtable}%
    \caption{Contract features of data for respective insurance types.}
\end{table}

\begin{table}[ht]
    \centering
    \begin{subtable}[c]{0.5\textwidth}
        \centering
        \begin{tabular}{l|c|c|c}
                \multicolumn{1}{c}{25th perc.} & \multicolumn{1}{c}{median}&  \multicolumn{1}{c}{75th perc.} &
                \multicolumn{1}{c}{max.}\\ \hline
                $471.44$ & $37~870.60$& $23~625.29$  & $785~665.97$  \\ \hline
        \end{tabular}
            \subcaption{Term Life Insurance.}
            \label{table_targets_TL}
    \end{subtable}%
    \begin{subtable}[c]{0.5\textwidth}
        \centering
        \begin{tabular}{l|c|c|c}
                \multicolumn{1}{c}{25th perc.} & \multicolumn{1}{c}{median}&  \multicolumn{1}{c}{75th perc.} &
                \multicolumn{1}{c}{max.}\\ \hline
                $330~492.43$ & $557~979.15$& $887~592.64$  & $3.52\cdot 10^6$  \\ \hline
        \end{tabular}
        \subcaption{Defined Contribution Plan.}
        \label{table_targets_pension}
    \end{subtable}%
    \caption{Statistics of maximal policy values $\max_t Y_t(\omega)$, see \eqref{eq_reserve_TL} or \eqref{eq_reserve_DC}, of simulated contracts $(X_1(\omega),\ldots,X_5(\omega))$.}
    \label{table_targets}
\end{table}

\paragraph{Term Life Insurance.} In a standard term life insurance contract the insurance company  pays the policyholder  a prespecified sum upon its death if the contract is still active. In our context, we uniquely identify a term life insurance contract by five features: the current age $X_1$ of the policyholder, the sum insured $X_2$, the duration of the contract $X_3$, the already lapsed fraction of the duration $X_4$ and the actuarial interest rate $X_5$ for discounting, which is influenced by a potential guaranteed interest of a contract. Each feature corresponds to a random variable $X_i$ with a bounded image space $X_i(\Omega):=X_i([0,1]^5)$, see Table \ref{table_features_values_TL}, from which we draw $N=100~000$ realizations. \\
The features sum insured $X_2$, duration $X_3$ and (actuarial) interest rate $X_5$ are obtained independently and uniformly, where $0.04$ is the maximal admissible actuarial interest rate for life insurance computations in Germany observed in history, see \cite{DAV_Hoechstrechnungszins}. Also, we restrict the duration (to $40$ years), since a term life insurance where the death of the policyholder occurs almost surely within the duration of the contract equals a whole life insurance, a different type of insurance. Next, the lapsed duration of a contract is a random, uniform fraction of its duration. This condition prevents matured contracts in our data. Lastly, for the current age $X_1$ we assume the initial age, at the start of the contract, to be uniformly distributed on $\lbrace 25,\ldots,67 \rbrace$, i.e. assuming no new contracts after the German age of retirement\footnote{To be precise, the age of retirement in Germany varies between 65 and 67 based on the date of birth. The basic retirement at 67 applies for individuals born on 1 January 1964 or later, see \cite{FED_Retirement_Germany}.}, which extends to the current age $X_1$ by adding the respective lapsed duration. Numerically, we utilize a $5$-dimensional Sobol sequence, which has been shown to cover higher-dimensional unit cubes rather uniformly, compare \cite{glasserman}, to get $N=100~000$ contracts and round their components to the nearest integer value, to avoid fractional durations. \\

In order to compute policy values, we make the following additional assumptions:
\begin{itemize}
    \item No expenses to the insurance company, no fractional durations of contracts and no lapses
    \item Constant sum insured, premium and interest rate, i.e. 
    no alterations to the setting of contracts throughout its duration, e.g. by profit repatriation
    \item Parametric survival model based on the Makeham Law, see \cite{dickson_hardy_waters_2009}, which results in a $t$-year survival probability ${}_t p_x$ of individual aged $x$ of
    \begin{align} \label{eq_SUSM}
        {}_t p_x = \exp\lbrace -\text{A}t-\frac{\text{B}}{\log(\text{c})}
                        \text{c}^x(\text{c}^t-1) \rbrace
    \end{align}
    In particular, we adopt the choice of \cite{dickson_hardy_waters_2009}, the SUSM model, by setting for the baseline hazard A$=0.00022$ and for age related factors B$=2.7\cdot 10^{-7}$ and c$=1.124$. In the Appendix in Figure \ref{fig_SUSM_DAV}, we provide an illustration of how this mortality assumption relates to mortality probabilities observed in Germany.
\end{itemize}

Based on the presented contract features and assumptions, we can calculate the expected policy values up to maturity  for an individual term life contract $X(\omega)=\left(X_1(\omega),\ldots,X_5(\omega)\right)$, $\omega\in[0,1]^5$, at times $t=0,\ldots,X_3(\omega)-1$ by
\begin{align} \label{eq_reserve_TL}
   &( {}_t V_{X_1(\omega)} + P(X(\omega)))(1+X_5(\omega)) =   
                    {}_1q_{X_1(\omega)+t}X_2(\omega)+
                    {}_1p_{X_1(\omega)+t}~{}_{t+1}V_{X_1(\omega)},
\end{align}
where the quantity $P(X(\omega)))$ represents the premium of the respective contract $X(\omega)$ based on the premium equivalence principle, see \cite{dickson_hardy_waters_2009}. As we assume no costs to the insurance company, for the initial policy value holds ${}_0 V_{X_1(\omega)} := 0$. \\
Lastly, in order to have target quantities $Y(\omega)=(Y_0(\omega),\ldots,Y_T(\omega))$ of equal length for all contracts, we zero-pad them and exclude past times, i.e.
\begin{align} \label{eq_zero_padding}
    Y_t(\omega)  =
   \begin{cases}
         {}_{t+X_4(\omega)} V_{X_1(\omega)} &\text{ for } t = 0,\ldots,X_3(\omega)-X_4(\omega), \\
        0 &\text{ for } t= X_3(\omega)-X_4(\omega)+1, \ldots, \max X_3(\Omega).
   \end{cases}
\end{align}

\paragraph{Defined Contribution Plan.} Pension plans are offered by employers to their the employees and provide financial support at retirement due to age, disability or potentially even benefits to the widow, resp. widower, in case of a premature death of the employee. The two most common types of pension planes are the defined benefits plan and defined contribution plan, see \cite{dickson_hardy_waters_2009, Neuburger19}. The first type has the terminal benefits to the employer set at the beginning of the plan, meaning the employers bears the financial risk of funding its obligations to the employee. The employee's contributions to the plan are therefore set to be in accordance with a risk neutral evaluation of the benefits. In contrast, the defined contribution plan has a fixed contribution, e.g. as a percentage of the employee's salary. The benefits are then a result of the employee's actual salary over time and the performance of the fund which the contributions were invested to. In practice, desirable contribution rates can be estimated by projecting the current salary and the fund value up to the age of retirement. Then, the contribution rate is set to meet a replacement rate, i.e. a percentage of the final salary which an annuitization of the expected final fund volume should pay throughout retirement. In a defined contribution plan, the risk of benefits being lower than expected is borne by the employee. The employer acts as a trustee, managing the employee's contribution and potentially granting subventions.\\
In our numeric analysis, we focus on the defined contribution type. Each defined contribution plan is assumed to be uniquely defined by five variables, the current age $X_1$ of the policyholder, the current fund volume $X_2$ (potentially including volume transferred from previous employments),  the current salary $X_3$, which can be projected to future years based on a constant growth factor, alias salary scale, $X_4$, and the contribution $X_5$ as a fixed share of the employee's salary. For the data of $N=100~000$ pension plans we utilize a 5-dimensional Sobol sequence and scale its realizations to the features' image space $X(\Omega)$, see Table \ref{table_features_pension}.\\

For defined contribution plans we assume the following:

\begin{itemize}
	\item No disability benefits, widow(er) benefits, lapses or suventions by the employer.
	\item Contributions are paid anually at the beginning of the year.
	\item A fixed, expected fund performance of $i=0.03$.
	\item In case of death prior to retirement the accumulated fund is payed out.
	\item Benefits, at death or retirement, are paid as lump sums at the end of the year. An annuitization of the lump sum equals a deferred whole life insurance, which is not part of this work and can be analyzed seperately.
	
	\item Retirement occurs, at the latest, at the age of $67$ or prematurely at ages $x = 60,\ldots,66$ with probabilities\footnote{The given probabilities are in line with the pension plan service table in \cite{dickson_hardy_waters_2009}, Table 10.2.} $(rr_{60},\ldots,rr_{66}) =  (0.3, 0.1, \ldots, 0.1)$ . For ages $x<60$ holds $rr_x=0$.
	\item Markov-type state transitions for states 'active', 'retired' and 'dead' are defined by the parametric mortality model from \eqref{eq_SUSM}, in combination with retirement rates $rr_x\in\left[0,1\right]$, i.e.
	\begin{multicols}{3}
    	\begin{itemize}
    		\item[a)] ${}_1p_x^{(\text{active})}=(1-rr_x)~ {}_1p_x$,
    		\item[b)] ${}_1q_x = (1-rr_x) (1-{}_1p_x)$,
    		\item[c)] ${}_1p_x^{(\text{retirement})}=rr_x$.
    	\end{itemize}
	\end{multicols}
\end{itemize}

Based on the given contract features and assumptions, we can compute the expected policy values, i.e. the expected fund volume, of a defined contribution plan at times $t=0,\ldots,67-X_1(\omega)$ by
\begin{align}\label{eq_reserve_DC}
        {}_{t} V_{X_1(\omega)} 
                    &  = \underbrace{{}_{t-1}V_{X_1(\omega)}(1+i)}_{\text{Expected Fund Growth}} + \underbrace{(1-rr_{X_1(\omega)+t-1})X_5(\omega) X_3(\omega)(1+X_4(\omega))^t(1+i)}_{\text{Expected Contribution}} \nonumber \\
                    & -\underbrace{rr_{X_1(\omega)+t-1}\cdot{}_{t-1}V_{X_1(\omega)}(1+i)}_{\text{Exp. Retirement Benefit, valued at $t$}}  \\
                     &-\underbrace{(1-rr_{X_1(\omega)+t-1})~{}_1q_{X_1(\omega)}\left({}_{t-1}V_{X_1(\omega)}+X_5(\omega) X_3(\omega)(1+X_4(\omega))^t\right)(1+i)}_{\text{Death Benefit, valued at $t$}} \nonumber \\
                    & = \left[ {}_{t-1}V_{X_1(\omega)} + X_5(\omega) X_3(\omega)(1+X_4(\omega))^t \right](1+i)(1-rr_{X_1(\omega)+t-1})~{}_1p_{X_1(\omega)} \nonumber
\intertext{with}
	    {}_0 V_{X_1(\omega)} &= X_2(\omega). \nonumber
\end{align}
Finally, we again apply zero-padding to ${}_tV_{X_2(\omega)}$  to obtain the target quantity $Y(\omega)=(Y_1(\omega),\ldots,Y_T(\omega))$ for DC plans, i.e.
$$Y_t(\omega)= \begin{cases}
         {}_t V_{X_1(\omega)} &\text{ for } t = 0,\ldots,67-X_1(\omega), \\
        0 &\text{ for } t= 67-X_1(\omega)+1, \ldots, 67-\min X_1(\Omega).
   \end{cases}$$


%% file: Sections/Section_Num_Results.tex
\section{Numerical Results} \label{sec_num_results}

Next, we provide numerical results for our methodology from Section \ref{sec_methodology} applied to data from Section \ref{sec_data}. We denoted term life and defined contribution contracts both by $X(\omega)$ and the respective policy values, i.e. target risk feature as defined in \eqref{eq_reserve_TL} and \eqref{eq_reserve_DC}, by $Y(\omega)=(Y_1(\omega),\ldots,Y_T(\omega))=R(X(\omega))$, where $\omega\in\Omega = [0,1]^5$. 

\paragraph{Prediction of Policy Values.}
    We recall, the task is to find a neural network $\hat{R}$, such that we minimize 
            $$\lVert \hat{R}(X(\omega)) - Y(\omega) \rVert, $$
    where $\omega\in\Omega$ is arbitrary and $\lVert\cdot\rVert$ a user-selected norm, see Section \ref{sec_grouping}. We provide and interpret results for both a mean, squared average (MSE), i.e. $\lVert x \rVert = \sum_{i=1}^{n}\tfrac{(x_i)^2}{n}$, as well as a mean, absolute average (MAE), i.e. $\lVert x \rVert = \sum_{i=1}^{n}\tfrac{\vert x_i \vert}{n}$. Hence, the training of our neural network is implemented either w.r.t. the mean squared error or the mean absolute error. \\
    
    Motivated by the recursive, time-dependent nature of the policy values, see \eqref{eq_reserve_TL} and \eqref{eq_reserve_DC}, we employ recurrent network structures as implemented in LSTM cells, respectively their GPU performance optimized CuDNNLSTM-version, see e.g. \cite{LSTM,chollet2015keras}. This intuitive choice is also backed by a preliminary analysis, where classical feed-forward networks showed a poor performance compared to recurrent structures of similar depth and number of parameters, see \cite{Kiermayer19}. In line with standard practice, we use scaled inputs, i.e. $\Tilde{\omega}:=2\omega-1\in[-1,1]^5,~\omega\in\Omega,$ instead of contract features $X(\omega)$. Lastly, the architecture includes a final scaling layer $\lambda$, resp. $\lambda^{\log}:~[-1,1]^T\rightarrow [\min_{t,\omega} Y_t(\omega)), \max_{t,\omega} Y_t(\omega) ]^T$, where
        \begin{align} \label{eq_lambda_layer}
            \lambda(z) &:= \frac{z+1}{2}(\max_{t,\omega} Y_t(\omega) - \min_{t,\omega} Y_t(\omega)), \\
            \lambda^{\log}(z) &:= \exp\left\lbrace\frac{z+1}{2}\log(1+\max_{t,\omega} Y_t(\omega) - \min_{t,\omega} Y_t(\omega))\right\rbrace -1 + \min_{t,\omega} Y_t(\omega).
        \end{align}
        
    If we recall the range of target values  $Y(\omega)=(Y_1(\omega),\ldots,Y_T(\omega)$ from Table \ref{table_targets}, we observe a high discrepancy between the range of scaled input values $\Tilde{\omega}\in[-1,1]^5$ and the range of targets $Y(\omega)\in\left[\min_{t,\omega}Y(\omega),\max_{t,\omega}Y(\omega) \right]$. This can lead to exploding weight parameters, which we avoid by employing a scaling layer $\lambda$, resp. $\lambda^{\log}$. By including the scaling layer we effectively train our model to approximate scaled targets $\lambda^{-1}(Y(\omega))\in\left[-1,1\right]^T$, resp. $\lambda^{-1}(Y(\omega))$. Since we compute the loss only after rescaling to target $Y(\omega)$, a gradient descent algorithm takes the actual losses into account and not a biased version. \\
    
\begin{figure}[htb]
        \centering
        \includegraphics[width=.6\textwidth]{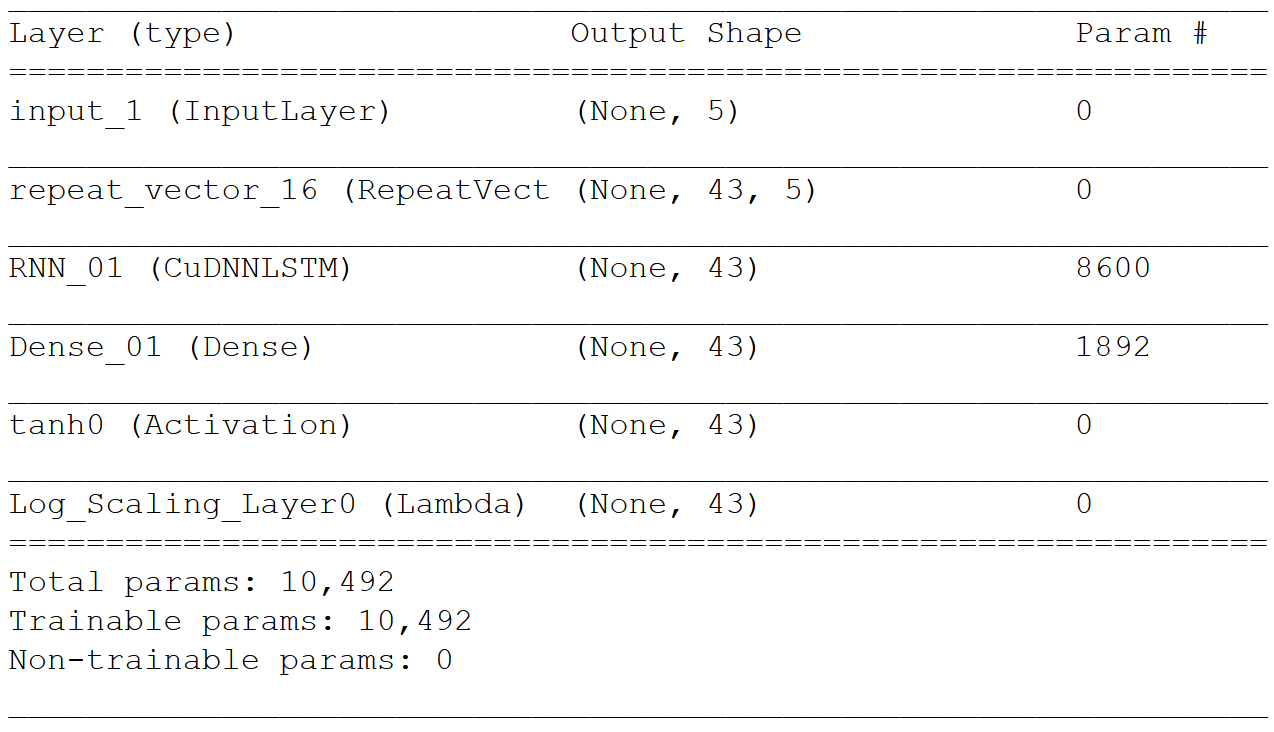}
        \label{fig_model_architecture_TL}
        \caption{Basic Model Architecture - Prediction of Policy Values of Term Life Insurance Contracts.}
        \label{fig_model_architectures}
    \end{figure}

    \begin{remark}
        We employ the $\lambda(\cdot)$ scaling for defined contribution plans and $\lambda^{\log}(\cdot)$ for term life insurance. For the later, the space of target values $Y$ is populated rather sparely, see Table \ref{table_targets_TL}, where the value of the 75th percentile is a mere $3\%$ of the maximum value. A logarithmic transformation provides a more evenly populated space $\log^{-1}(Y(\Omega))$ for training, see Appendix Figure \ref{fig_target_scaling}.
    \end{remark}
    Figure \ref{fig_model_architectures} provides a summary of the architecture of a single neural network and its components for predicting the policy values of defined contribution plans. To boost performance, we will eventually employ ensembles of networks of equal architecture, but different parameters, e.g. due to stochasticity of their training or different loss functions. Additionally to control for overfitting, we use early stopping w.r.t. the validation loss, see \cite{HaDeBeJe}, and train models for a maximum of $N_{\text{Ens}}=1500$ epochs. The training process itself is performed by an adaptive moment estimation algorithm, i.e. 'adam' and its standard parameters in term of learning rate and decay rates, see e.g. \cite{adam, chollet2015keras}.

    Next, we analyse the level of accuracy, with which neural networks can replicate standard, actuarial computations $R(\cdot)$ of policy values of term life insurances and defined contribution plans, see \eqref{eq_reserve_TL} and \eqref{eq_reserve_DC}. In line with standard practice, see \cite{HaDeBeJe}, we split our data into test set $\mathcal{D}_{test}$ ($30\%$ of data) and training set $\mathcal{D}_{train}$ ($70\%$ of data), of which we take $25\%$ for out-of sample validation and to determine the time of early stopping. We present results of ensemble models, with sub-models trained w.r.t. mean squared error or mean absolute error, for the respective contract types separately. In order to evaluate the performance of $\hat{R}$ for a contract $x\in\mathbb{R}^5$ at time $t\in\lbrace 0,\ldots,T\rbrace$, we consult the absolute error $e_t$ and relative error re${}_t$, if $R(x)_t>0$, given by
    \begin{align} \label{eq_error_types}
        e_t(x) &:= \hat{R}(x)_t-R(x)_t,  &
        \text{re}{}_t(x) &:= \frac{\hat{R}(x)_t-R(x)_t}{R(x)_t}.
    \end{align}
    Lastly, we are also interested in the relative error of an individual prediction $\hat{R}(x)$ from a portfolio point of view. For example, in practice the severity of a relative error re${}_t(x)=2$ depends on the volume of the respective policy value $R(x)$ and how it compares to other policy values $R(x^{\prime})$, $x^{\prime}\in P$. The combination of $\hat{R}(x)=2$, with $R(x)=1$ is in general considered to be less severe than $\hat{R}(x)=2\cdot 10^5$, with $R(x)=10^5$. Hence, we introduce the weighted relative error wre${}_t$ of contract $x\in P$ relative to some portfolio $P$, e.g. $P=\mathcal{D}_{test}$, by
    \begin{align}
        \text{wre}_t(x;P) &:= \frac{\hat{R}(x)_t-R(x)_t}{\sum_{x^{\prime}\in P}R(x^{\prime})_t}.\label{eq_wre}
    \end{align}

    \paragraph{a) Term Life Insurance.\\}
    Table \ref{table_pred_TL_comparison} shows various models, whose architecture was discussed in the previous paragraph, with varying loss functions and ensemble sizes $N_{\text{Ens}}$, i.e. the number of individual models included in an ensemble. In addition to the mean value of error measures $e_t(x)$ and wre${}_t(x;\mathcal{D}_{\text{test}})$ for $x\in\mathcal{D}_{\text{test}}$, i.e. $\overline{\text{e}}_t$ and $\overline{\text{wre}}_t$, we also consider the empirical $99$th percentile $\text{pc}_{0.99,\vert\cdot\vert}$ of their absolute values, which indicate the variability of the respective error.

        \begin{table}[htb]
            \centering
            \includegraphics[width=.7\textwidth]{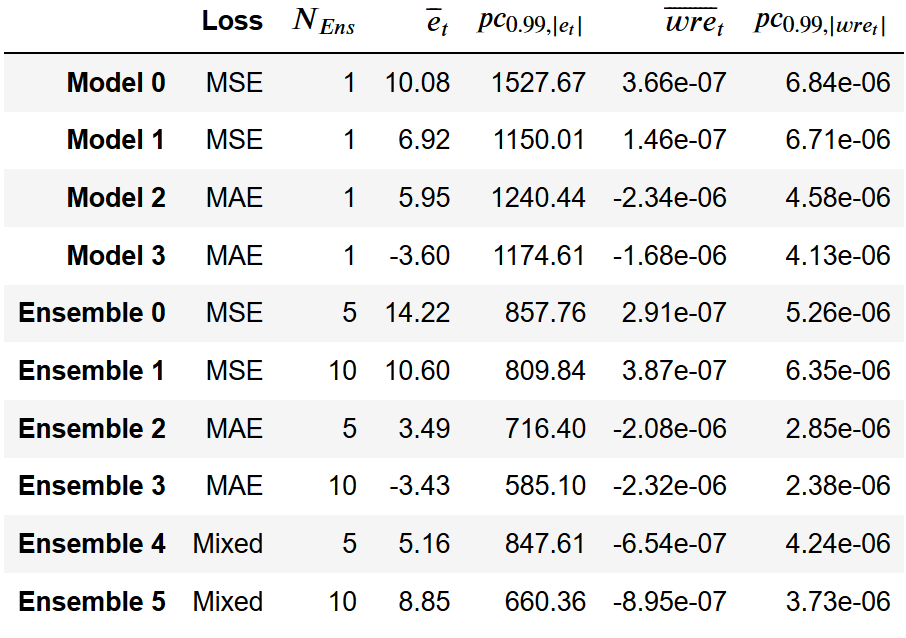}
            \caption{Comparison of prediction models for term life insurance contracts $x\in\mathcal{D}_{\text{test}}$. A Mixed loss includes MSE and MAE trained models.}
            \label{table_pred_TL_comparison}
        \end{table}
    
    Overall, errors $e_t$ indicate very little bias of our models. Yet we see the stochasticity in the training procedure, compare e.g. Model 2, with an average overshoot of $\overline{\text{e}}_t=5.95$, to Model 3, with an average shortcoming of $\overline{\text{e}}_t=-3.60$. In general, MAE training results in a lower mean error $\overline{\text{e}}_t$ than MSE training, which is reasonable as the MAE minimizes (absolute) differences. Also, we observe low weighted relative errors wre${}_t$ for all models, where training w.r.t. a MSE loss function dominates MAE training in terms of the mean value $\overline{\text{wre}}_t$. 
    While ensemble models are not guaranteed to improve all statistics, compare e.g. $\overline{\text{e}}_t$ for Model 0 and Ensemble 0, we see a reduction in variability, resp. the 99th percentile $\text{pc}_{0.99,\vert\cdot\vert}$, which is a basic motivation for applying ensemble models. In fact, ensemble methods work best if the sub-models show little correlation, see \cite{GoBeCo}. Hence, we also include mixed ensembles, which contain sub-models trained w.r.t. MSE and MAE, however, with no clear indication of superiority. \\
    Note that for the task of grouping, where a cluster $C$ is replaced by a model point $\Tilde{x}$, the accuracy of $\hat{R}(\Tilde{x})$ dictates the quality of the grouping for the full cluster $C$. Hence, we also look at the relative error re${}_t(x)$ for individual contracts $x\in\mathcal{D}_{\text{test}}$ and present results subdivided w.r.t. their target values relative to the overall maximal target value, i.e. $\frac{\max_t R(x)}{\max_{x^{\prime}, t} R(x^{\prime})}\in\left[0,1\right]$. Figure \ref{table_re_wrt_volume} shows that relative errors re${}_t$ for MSE training are in general quite low, however, contracts with low policy values can be replicated poorly. This is an intuitive consequence of the quadratic MSE, which emphasises large values. An increase of the number of sub-models included in an ensemble can mitigate this flaw, compare e.g. Tables \ref{table_re_mse_5} and \ref{table_re_mse_10}.\\
        \begin{table}[htb]
            \centering
            \begin{subtable}{.8\textwidth} 
                \centering
                \includegraphics[width=\textwidth]{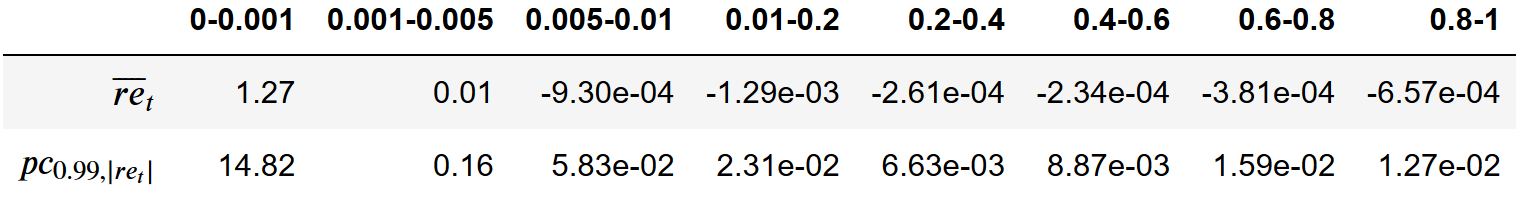}
                \subcaption{Ensemble 0, i.e. $N_{\text{Ens}}=5$, MSE-Loss.}
                \label{table_re_mse_5}
            \end{subtable}
            \begin{subtable}{.8\textwidth} 
                \centering
                \includegraphics[width=\textwidth]{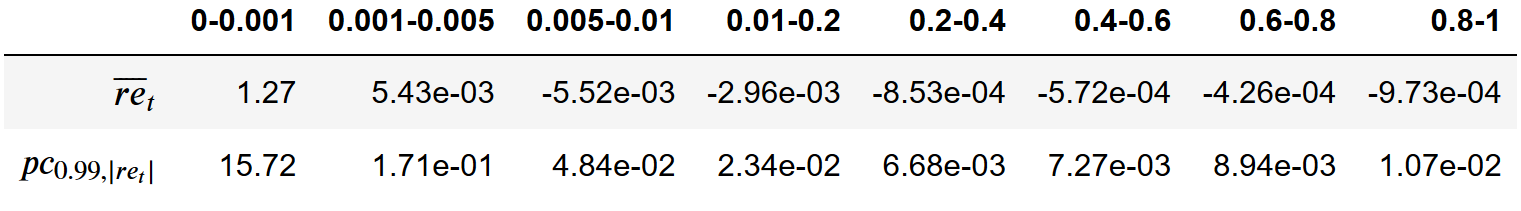}
                \subcaption{Ensemble 1, i.e. $N_{\text{Ens}}=10$, MSE-Loss.}
                \label{table_re_mse_10}
            \end{subtable}%
            \caption{Relative error re$_t(x)$ for term life contracts $x\in\mathcal{D}_{\text{test}}$, split w.r.t. $\frac{\max_t R(x)}{\max_{x^{\prime}, t} R(x^{\prime})}\in\left[0,1\right]$. }
            \label{table_re_wrt_volume}
        \end{table}
        
    For the subsequent grouping procedure of a term life insurance portfolio, we set $\hat{R}$ to be Ensemble 0, i.e. an ensemble of $5$ sub-models, each trained w.r.t. a MSE loss. Although the performance of Ensemble 0 might still be improved, with additional computational effort, Ensemble 0 will prove to be sufficient to provide significant improvements of a standard $K$-means grouping.
    \newpage
    \paragraph{b) Defined Contribution Plan.\\}
    Similar to term life insurance, we analyse various model architectures $\hat{R}$ for defined contribution plans and find low weighted relative errors wre${}_t(x;\mathcal{D}_{\text{test}})$, as well as low errors e${}_t(x)$ for pension plans $x\in\mathcal{D}_{\text{test}}$. Comprehensive results for $e_t$ and wre${}_t$ are provided in the Appendix, see Table \ref{table_comparision_prediction_pension}. Note that the portfolio of defined contribution plans exhibits significantly higher policy values than the term life insurance portfolio, see Table \ref{table_targets}. Hence, errors e${}_t$, which are higher than in the previous section for term life insurance, do not correspond to poor training. \\
    Table \ref{table_pensions_re_wrt_volume} presents a low relative error re${}_t(x)$ for pension plans $x\in\mathcal{D}_{\text{test}}$ for most parts. However, we again observe a significant increase of re${}_t(x)$ for MSE trained models and plans with low policy values. This flaw can be mitigated by increasing $N_{\text{Ens}}$ or using a MAE training, which adjusts the re${}_t$ more evenly for different target volumes, see Table \ref{table_pension_re_mae_5}.

    \begin{table}[htb]
        \centering
        \begin{subtable}{.7\textwidth} 
            \centering
            \includegraphics[width=\textwidth]{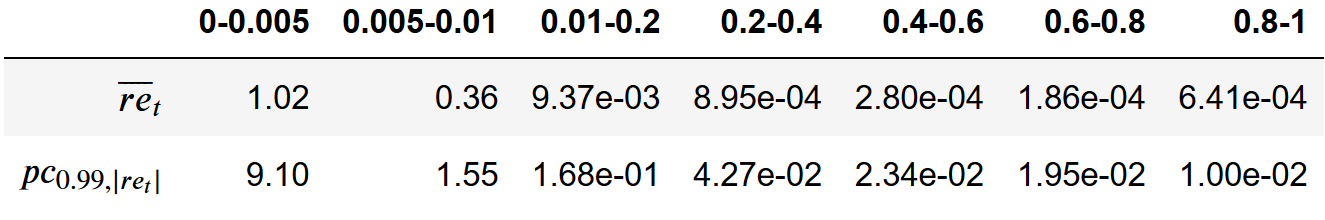}
            \subcaption{Ensemble 0, i.e. $N_{\text{Ens}}=5$, MSE-Loss.}
            \label{table_pension_re_mse_5}
        \end{subtable}
        \begin{subtable}{.7\textwidth} 
            \centering
            \includegraphics[width=\textwidth]{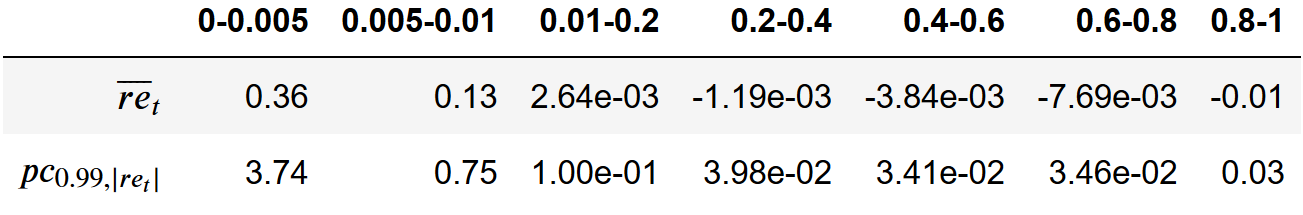}
            \subcaption{Ensemble 2, i.e. $N_{\text{Ens}}=5$, MAE-Loss.}
            \label{table_pension_re_mae_5}
        \end{subtable}%
        \caption{Relative error re$_t(x)$ for defined contribution plans $x\in\mathcal{D}_{\text{test}}$, split w.r.t. $\frac{\max_t R(x)}{\max_{x^{\prime}, t} R(x^{\prime})}\in\left[0,1\right]$. }
        \label{table_pensions_re_wrt_volume}
    \end{table}
        
    To avoid poor accuracies for clusters of contracts with low policies values, see Table \ref{table_pension_re_mse_5} with $\overline{\text{re}}_t(x)=0.36$ for pension plans $x\in\mathcal{D}_{\text{test}}$ and $\frac{\max_t R(x)}{\max_{x^{\prime}, t} R(x^{\prime})}\in \left[0.005,0.01\right]$, we favour Ensemble 2 over Ensemble 0, i.e. choose an ensemble of $5$ sub-models trained w.r.t. MAE.

\paragraph{Grouping of Portfolios.} Next, we apply the grouping methodology from Section \ref{sec_methodology}, using prediction models discussed in previous paragraphs, and compare it to a $K$-means baseline, where model points are equal to centroids. 
The portfolio to be grouped is the complete data simulated in Section \ref{sec_data}, i.e. including $\mathcal{D}_{\text{train}}$ and $\mathcal{D}_{\text{test}}$. For the optimization we use the 'Adadelta' algorithm without an adaptive learning rate, see \cite{chollet2015keras}, since we want to search the space $\Tilde{\Omega}=[-1,1]^5$ of scaled contract features $X_1,\ldots,X_5$ preferably evenly. The loss function of the larger network for grouping, with nested and fixed $\hat{R}$, is the MSE loss. \\ 
Note, that the original computation of target values $R(P)$ was performed w.r.t. integer-valued current age, duration and lapsed duration, see \eqref{eq_reserve_TL} and \eqref{eq_reserve_DC}. As in general neither $K$-means centroids $\Tilde{c}$, nor the optimized centroids $\Tilde{x}$ are integer valued after rescaling them to the range of $X=(X_1,\ldots,X_5)$, we use approximate upper and lower bounds to compare the classical computation, i.e. $R(\Tilde{c})$ and $R(\Tilde{x})$, with the model prediction $R(\Tilde{x})$. For term life we floor the duration $X_3$ and ceil the lapsed duration $X_4$, and vice versa, to get lower and upper bound. For pension plans, we floor and ceil the current age $X_1$, which determines the time to retirement, to get bounds. Also, we set the $K$-means prediction $R(\Tilde{x})$ equal to the mean of upper and lower bound.
    
    \paragraph{a) Term Life Insurance.\\}
    
\begin{figure}[htb]
        \centering
        \begin{subfigure}{.4\textwidth} 
            \centering
            \includegraphics[width=.9\textwidth]{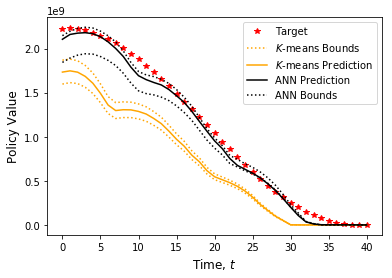}
            \subcaption{$K=100$.}
            \label{fig_grouping_K100}
        \end{subfigure}
        \begin{subfigure}{.4\textwidth} 
            \centering
            \includegraphics[width=.9\textwidth]{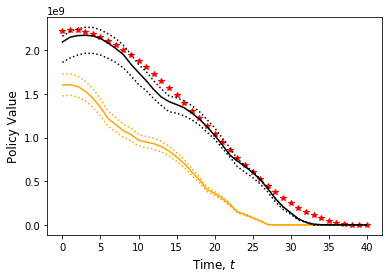}
            \subcaption{$K=50$.}
            \label{fig_grouping_K50}
        \end{subfigure}
        \begin{subfigure}{.4\textwidth} 
            \centering
            \includegraphics[width=.9\textwidth]{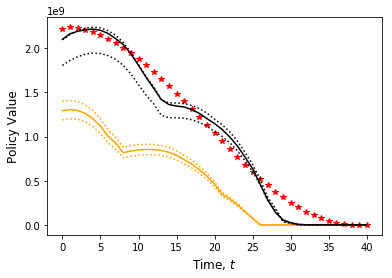}
            \subcaption{$K=25$.}
            \label{fig_grouping_K25}
        \end{subfigure}
        \begin{subfigure}{.4\textwidth} 
            \centering
            \includegraphics[width=.9\textwidth]{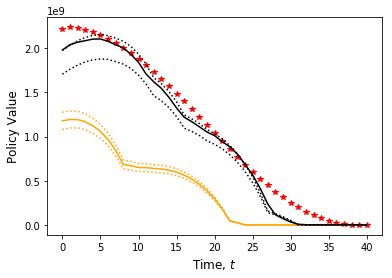}
            \subcaption{$K=10$.}
            \label{fig_grouping_K10}
        \end{subfigure}
        \caption{Grouping of term life insurance portfolio with $N=100,000$ contracts to $K$ model points.}
        \label{fig_grouping_TL}
    \end{figure}

    Figure \ref{fig_grouping_TL} illustrates the quality of grouping for various degrees of compression, i.e. reducing the number of model points from $N=100~00$ in portfolio $P$ to $K\in\lbrace10,25,50,100\rbrace$ model points in grouping $\Tilde{P}$. In general, we see that the optimized model point approach using a neural network (ANN) and its prediction $\hat{R}(\Tilde{P})$ outperform $K$-means, i.e. grouping $P_{\text{KM}}$ and risk feature $R(P_{\text{KM}})$. In fact, while the quality of grouping via a neural network remains relatively stable for a lower number of model points $K$, the performance of $K$-means deteriorates significantly as $K$ decreases and clusters become more heterogeneous. In terms of statistical measures, the ANN prediction $\hat{R}(\Tilde{P})$ outperforms $R(P_{\text{KM}})$ w.r.t. the mean, relative error $\overline{\text{re}}_t$ by a factor of close to $3$. For details see Table \ref{table_grouping_TL} in the Appendix. Similarly, the mean, absolute error $\overline{\text{e}}_t$ of $\hat{R}(\Tilde{P})$ compared to $R(P_{\text{KM}})$ drops by a factor in the range of $7$ to $13$. The nature of those improvements also holds true if we backtest with approximate lower and upper bounds, which apply the traditional computation scheme $R(\cdot)$. 
    
    Note that model points are optimized per cluster and the respective results are aggregated to obtain the policy values of the entire portfolio. In  the Appendix, Figure \ref{fig_K100_selected}, we illustrate the improvement of using the neural network approach for $K=100$ and two representative clusters. \\
    Overall, numerical results suggest a significant improvement over a $K$-means baseline for a given number of clusters $K$. The stabilized quality of grouping by the supervised learning concept allows for higher degrees of compression, i.e. lower $K$.
 
    \paragraph{b) Defined Contribution Plan. \\}
    Next, we repeat the procedure for defined contribution plans. Results for $K=10$ model points are provided in Figure \ref{figure_pens_grouping_K10} and its statistics in the Appendix, see Table \ref{table_pens_grouping_K10}. From the previous selection of $K\in\lbrace 10,25,50,100\rbrace$, $K=10$ is the highest degree of compression and supervision of the grouping should provide the most benefit, as preprocessed clusters are more heterogeneous than for higher values of $K$. Again, we find the ANN predicion to improve the $K$-means clustering in terms of mean relative error $\overline{\text{re}}_t$ and mean error $\overline{\text{e}}_t$, see Appendix Table \ref{table_pens_grouping_K10}. However, if we respect the approximate bounds of policy values of the optimized model points, we do not find the neural network approach to significantly outperform grouping based on $K$-means clustering. \\
    Note that centroids of $K$-means clustering provide optimal model points if the target quantity is linear w.r.t. all contract features. Recalling \eqref{eq_reserve_DC}, we realize that in our setting policy values of defined contribution plans are perfectly linear w.r.t. the initial fund volume $X_2$, salary $X_3$ and contribution rate $X_5$. Only the initial age $X_1$ and salary scale $X_4$ introduce non-linearity. Retirement rates at ages $x\in\lbrace 60,\ldots,67\rbrace$ are exogenous. Hence, the very similar results of the $K$-means baseline and the neural network approach are of little surprise. So far, the grouping procedure was restricted to represent the mixture of policy values from multiple pension plans by the policy values of a single pension plan. Hence, a natural generalization is to allow for multiple model points per cluster, as discussed in Remark \ref{rem_grouping_method}. An illustration of the effect of mixing policy values of multiple model points in a single cluster can be found in the Appendix, see Figures \ref{fig_pension_grouping_K10_1_selected} and \ref{fig_pension_grouping_K10_2_selected}. \\
    
    \begin{figure}[htb]
        \centering
        \includegraphics[width=.5\textwidth]{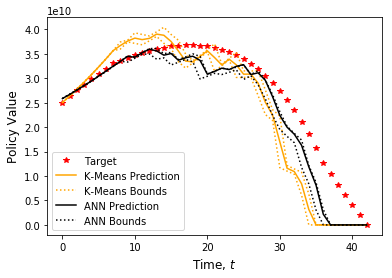}
        \caption{Grouping of portfolio of $N=100,000$ defined contribution plans and $K=10$ model points.}
        \label{figure_pens_grouping_K10}
    \end{figure}
    
\begin{figure}[htb]
        \centering
        \begin{subfigure}{.49\textwidth} 
            \centering
            \includegraphics[width=.9\textwidth]{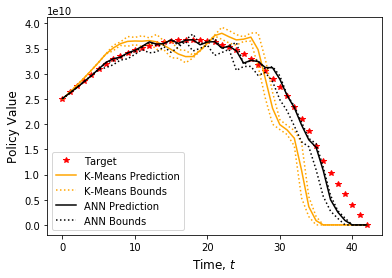}
            \subcaption{$25$-means clustering vs. ANN approach with $5$-means preprocessing, with $5$ model points per cluster.}
            \label{fig_grouping_pension_K5_C5}
        \end{subfigure}
        \begin{subfigure}{.49\textwidth} 
            \centering
            \includegraphics[width=.9\textwidth]{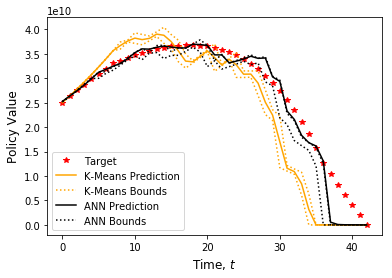}
            \subcaption{$10$-means clustering vs. ANN approach with $1$-means preprocessing, with $10$ model points per cluster.}
            \label{fig_grouping_pension_K1_C10}
        \end{subfigure}
        \caption{Comparison of $K$-means grouping versus ANN approach with multiple model points per cluster.}
        \label{fig_grouping_pension_multiple_mps}
    \end{figure}

    Finally, we compare performances of the $K$-means baseline to the neural network approach with multiple model points per cluster, where both methods utilize equally many model points. All model points of the same cluster are set to have the same relevance, i.e. we have output $o=\tfrac{1}{m}\sum_{i=1}^{m}\hat{R}(\Tilde{x}^{(i)})$ as discussed in Remark \ref{rem_grouping_method}. For illustration, we look at a $K$-means grouping with $K\in\lbrace 10,25  \rbrace$ and compare it to our ANN approach with $K$-mean preprocessed portfolio, $K\in\lbrace 1, 5 \rbrace$, where each cluster is assigned $10$, resp. $5$ model points, see Figure \ref{fig_grouping_pension_multiple_mps}. Note, that for $K=1$ we effectively remove the requirement of processing the portfolio by some clustering procedure. The generalization of multiple model points per cluster shows significant improvements in replicating the policy values of the portfolio more smoothly, especially close to maturity, when only pension plans with low volume policy values remain. We present statistical results in Table \ref{table_grouping_pension_multiple_mps}. Compared to $K$-means, the ANN prediction lessens the mean relative error $\overline{\text{re}}_t$ from $-0.21$, resp. $-0.24$, to $-0.10$, resp. $-0.12$, which is a reduction by the rough factor of $2$. Similarly, we see a big drop of the mean error $\overline{\text{e}}_t$ by the factor of roughly $3$ to $4$ when by apply the ANN approach instead of $K$-means. Note that for each time $t$, $\text{e}_t$ is the aggregated error for a portfolio of $N=100~000$ pension plans. The quality of those results holds still true if we consider approximate bounds for the policy values of ANN model points.

    \begin{table}[htb]
        \centering
        \begin{subtable}{.4\textwidth} 
            \centering
            \includegraphics[width=.9\textwidth]{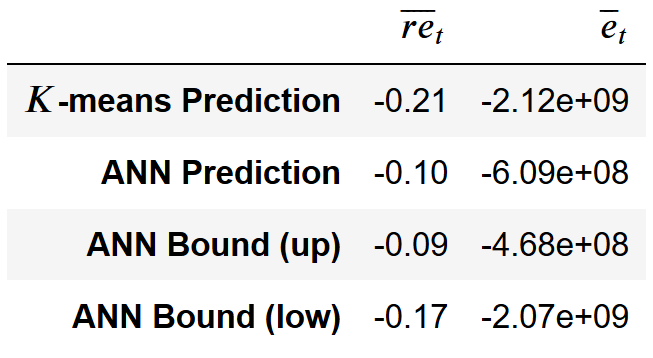}
            \subcaption{Statistics of Figure \ref{fig_grouping_pension_K5_C5}.}
            \label{table_grouping_pension_K5_5}
        \end{subtable}
        \begin{subtable}{.4\textwidth} 
            \centering
            \includegraphics[width=.9\textwidth]{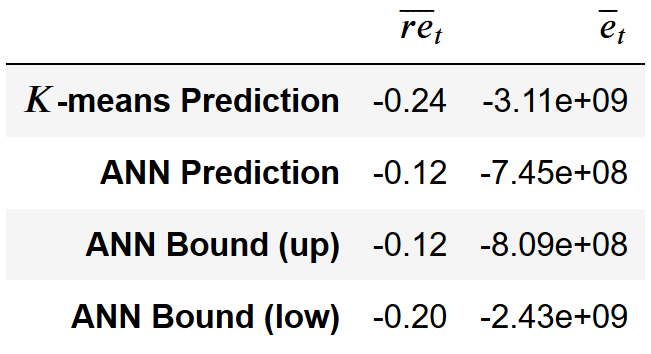}
            \subcaption{Statistics of Figure \ref{fig_grouping_pension_K1_C10}.}
            \label{table_grouping_pension_K1_C10}
        \end{subtable}
        \caption{Comparison of $K$-means grouping versus ANN approach with multiple model points per cluster.}
        \label{table_grouping_pension_multiple_mps}
    \end{table}

    \newpage

%% file: Sections/Section_Conclusion.tex
\section{Conclusion} \label{sec_conclusion}
Our novel concept for grouping of contracts in insurance utilizes neural networks and includes a supervision of its quality. The quality of a proposed grouping is evaluated for selected risk features $R_1,\ldots,R_l$ and w.r.t. user-selected norms. In order to supervise risk features, we first construct a neural network, i.e. a prediction model, which is capable of replicating the traditional computation of the respective risk features $R_1,\ldots,R_l$. Next, we nest this prediction model in a larger model and apply backpropagation to find optimal model points, which substitute preprocessed clusters of the respective portfolio and preserve features $R_1,\ldots,R_l$. \\
We provide numerical results for simulated portfolios of term life insurance contracts, as well as defined contribution pension plans. In this context, we consider policy values up to maturity as a single risk features of interest. After building adequate prediction models for policy values, we use $K$-means clustering to preprocess the respective portfolios and compare the performance of model points proposed by our approach with $K$-means centroids. We find significant improvements for grouping in term life insurance and minor improvements for defined contribution plans, which is explained by a linear relation of the target quantity w.r.t. a majority of the contract features. We further generalize the concept by substituting a single cluster of contracts by multiple model point. This generalization leads to significant improvements for grouping of defined contribution plans and enables us to remove the requirement of preliminary clustering of the portfolio altogether. \\
A critical question of grouping is the optimal number $K$ of model points to use. So far, when we allowed for multiple model points per cluster, we set them to have equal importance. Future research should analyze our methodology, also without requiring a preliminary clustering, and allow for varying importance of different model points. Additional weight regularization of model points, similar to e.g. Lasso regression, seems promising to numerically determine an optimal number of clusters $K$. We also encourage future research to investigate the inclusion of qualitative contract features and the application of dropout, to potentially and efficiently replace ensemble structures.

%% file: Sections/Appendix.tex
\appendix
\section{Appendix}
\subsection{Figures}

    \begin{figure}[H]
        \centering
        \includegraphics[width=.7\textwidth]{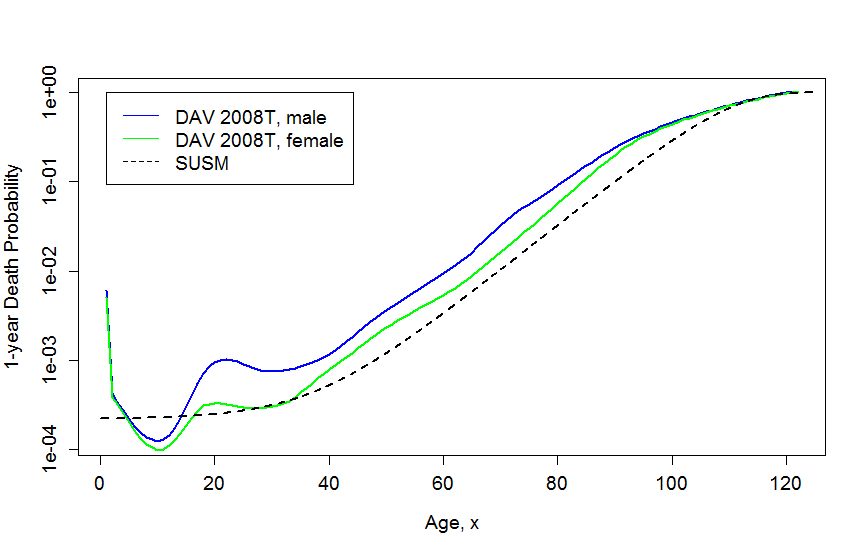}
        \caption{Comparison of single year death probabilities.}
        \label{fig_SUSM_DAV}
    \end{figure}

    \begin{figure}[H]
        \centering
        \includegraphics[width=.9\textwidth]{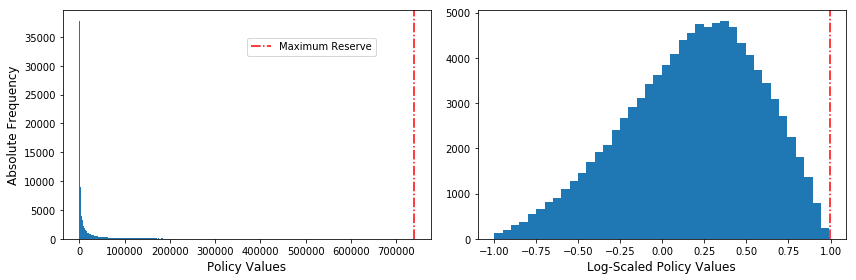}
        \caption{Logarithmic target scaling for term life insurance portfolio.}
        \label{fig_target_scaling}
    \end{figure}
    
    \begin{figure}[H]
        \centering
        \includegraphics[width=.9\textwidth]{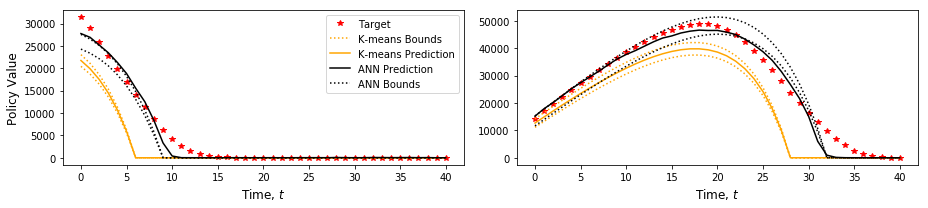}
        \caption{Grouping of term life portfolio, $K = 100$ clusters, each with $1$ model point. Illustration of representative clusters.}
        \label{fig_K100_selected}
    \end{figure}

    \begin{figure}[H]
        \centering
        \includegraphics[width=.9\textwidth]{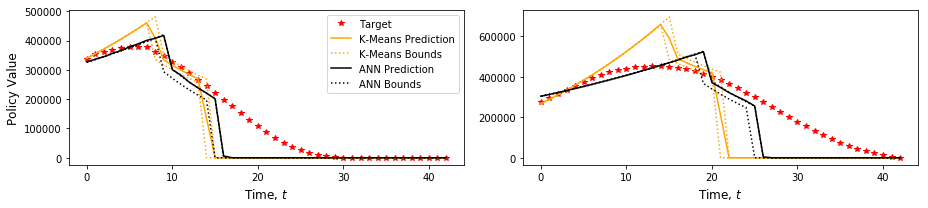}
        \caption{Grouping of pension plan portfolio, split into $K = 10$ clusters by the $K$-means algorithm. Each cluster is replaced by one model point. Two representative clusters are shown.}
        \label{fig_pension_grouping_K10_1_selected}
    \end{figure}

    \begin{figure}[H]
        \centering
        \includegraphics[width=.9\textwidth]{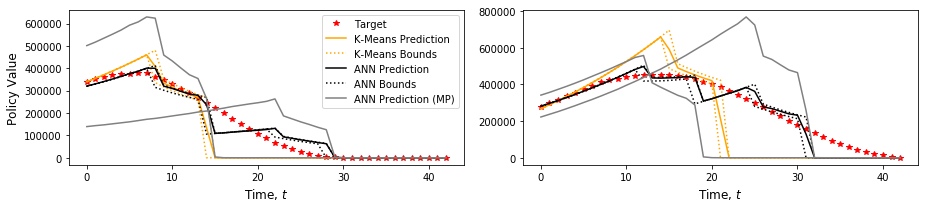}
        \caption{Grouping of pension plan portfolio, split into $K = 10$ clusters by the $K$-means algorithm. Each cluster is replaced by two model points (MP). The two representative clusters from Figure \ref{fig_pension_grouping_K10_1_selected} are revisited.}
        \label{fig_pension_grouping_K10_2_selected}
    \end{figure}

\subsection{Tables}

\begin{table}[H]
    \centering
    \includegraphics[width=.7\textwidth]{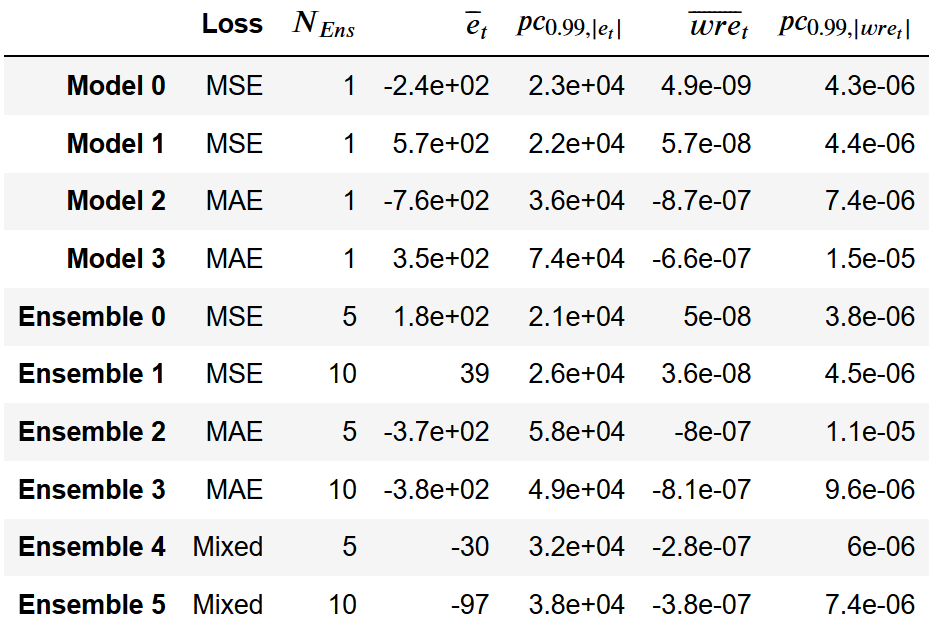}
    \caption{Comparison of prediction models for defined contribution plans $x\in\mathcal{D}_{\text{test}}$. A Mixed loss includes MSE and MAE trained models.}
    \label{table_comparision_prediction_pension}
\end{table}

\begin{table}[H]
    \centering
    \begin{subfigure}{.4\textwidth} 
        \centering
        \includegraphics[width=.9\textwidth]{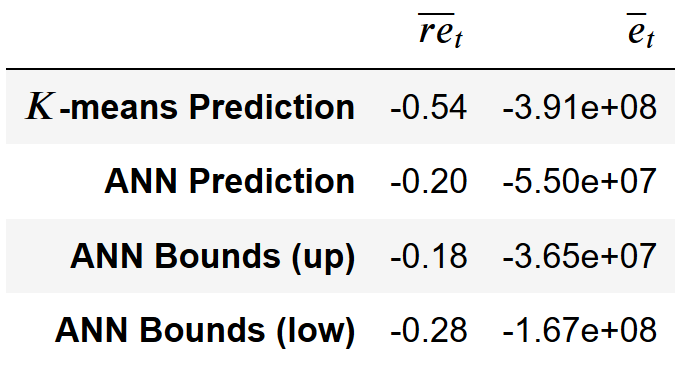}
        \subcaption{$K=100$.}
        \label{fig_grouping_K100_stat}
    \end{subfigure}
    \begin{subfigure}{.4\textwidth} 
        \centering
        \includegraphics[width=.9\textwidth]{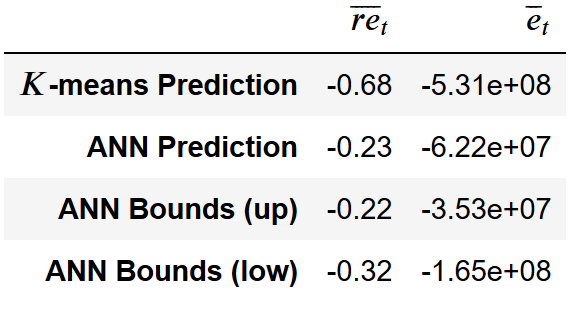}
        \subcaption{$K=50$.}
        \label{fig_grouping_K50_stat}
    \end{subfigure}
    \caption{Grouping of term life insurance portfolio with $N=100~000$ contracts to $K$ model points.}
    \begin{subfigure}{.4\textwidth} 
        \centering
        \includegraphics[width=.9\textwidth]{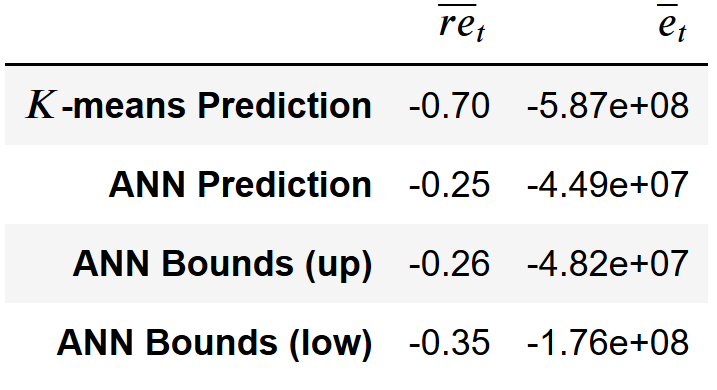}
        \subcaption{$K=25$.}
        \label{fig_grouping_K25_stat}
    \end{subfigure}
    \begin{subfigure}{.4\textwidth} 
        \centering
        \includegraphics[width=.9\textwidth]{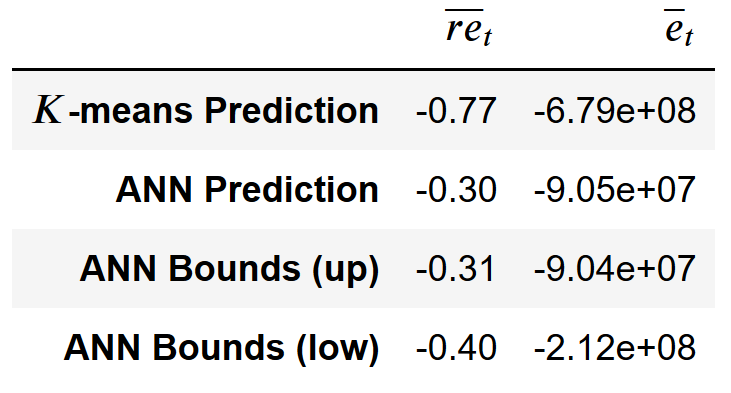}
        \subcaption{$K=10$.}
        \label{fig_grouping_K10_stat}
    \end{subfigure}
    \caption{Grouping of term life insurance portfolio with $N=100~000$ contracts to $K$ model points.}
    \label{table_grouping_TL}
\end{table}

\begin{table}[H]
    \centering
    \includegraphics[width=.36\textwidth]{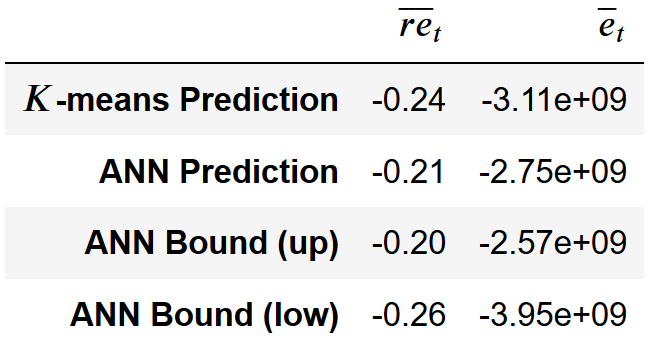}
    \caption{Statistics for Figure \ref{figure_pens_grouping_K10}. Grouping of pension portfolio with $N=100~000$ and $K=10$.}
    \label{table_pens_grouping_K10}
\end{table}